\documentclass[letterpaper, 10 pt, conference]{IEEEtran}

\usepackage[utf8]{inputenc}
\usepackage[T1]{fontenc}
\usepackage{graphicx}
\usepackage{subcaption}
\usepackage{amsmath}
\usepackage{amssymb}

\usepackage[normalem]{ulem}

\title{
On the Deployability of Augmented Reality Using Embedded Edge Devices
}

\author{
    \IEEEauthorblockN{Ayoub Ben Ameur\textsuperscript{\textsection}, Andrea Araldo}
    \IEEEauthorblockA{Télécom SudParis - Institut Polytechnique de Paris\\
    \{first\_name\}.\{last\_name\}@telecom-sudparis.com}
    \and
    \IEEEauthorblockN{Francesco Bronzino\textsuperscript{\textsection}}
    \IEEEauthorblockA{Université Savoie Mont Blanc\\
    francesco.bronzino@univ-smb.fr}
}

\usepackage{array}
\newcolumntype{C}[1]{>{\centering\arraybackslash}m{#1}}
\usepackage{pgfplots}
\pgfplotsset{width=8cm,compat=1.9}
\begin{document}

\maketitle
\thispagestyle{plain}
\pagestyle{plain}

\begingroup\renewcommand\thefootnote{\textsection}
\footnotetext{Work done while at Nokia Bell Labs France}
\endgroup

%%%%%%%%%%%%%%%%%%%%%%%%%%%%%%%%%%%%%%%%%%%%%%%%%%%%%%%%%%%%%%%%%%%%%%%%%%%%%%%%
\begin{abstract}

Edge Computing exploits computational capabilities deployed at the very edge of
the network to support applications with low latency requirements. Such
capabilities can reside in small embedded devices that integrate dedicated
hardware -- e.g., a GPU -- in a low cost package. But these devices have limited computing capabilities compared to standard server grade equipment.
When deploying an Edge Computing based application, understanding whether the available hardware can meet target requirements is key in meeting the expected performance.
In this paper, we study the feasibility of deploying Augmented Reality
applications using Embedded Edge Devices (EEDs). We compare such deployment
approach to one exploiting a standard dedicated server grade machine. Starting
from an empirical evaluation of the capabilities of these devices, we propose a
simple theoretical model to compare the performance of the two approaches. We
then validate such model with NS-3 simulations and study their feasibility. 
Our results show that there is no one-fits-all solution. If we need to deploy high responsiveness applications, we need a centralized server grade architecture and we can in any case only support very few users. The centralized architecture fails to serve a larger number of users, even when low to mid responsiveness is required. In this case, we need to resort instead to a distributed deployment based on EEDs.
\end{abstract}

%%%%%%%%%%%%%%%%%%%%%%%%%%%%%%%%%%%%%%%%%%%%%%%%%%%%%%%%%%%%%%%%%%%%%%%%%%%%%%%%

% \section*{Keywords}
% Edge Computing, Artificial Intelligence, Augmented Reality, Real-time Object Detection, Video Processing 

\section{Introduction}

% First paragraph on new applications like AR that require computing and low latency
The massive growth in popularity of real time applications, e.g., Augmented Reality (AR) and the Internet of Things (IoT), has pushed service providers to part ways from the standard Cloud Computing paradigm and move towards decentralized solutions like Mobile Edge Clouds (MEC) \cite{mec}. MECs are more distributed and localized forms of cloud computing that aim to minimize the service response time of deployed applications by offering computing capabilities at the edge of the network instead that from a centralized remote location. Applications benefit from the processing computing advantages of cloud computing without compromising on their real time requirements.

% Second paragraph on the heterogeneity and the intrinsic challenges that causes
To further carry forward the adoption of MECs, hardware providers have started to develop tiny embedded devices that integrate dedicated computing capabilities, e.g. a GPU or a TPU, at a reduced cost. NVIDIA's Jetson Nano \cite{jetson} and Google's Coral Dev Unit \cite{coral} are just two examples of this trend, providing a good balance between cost and performance. This is in contrast to more expensive approaches that require the deployment of expensive server grade hardware. On the other hand, adopting single board devices for computing intensive applications can introduce a new bottleneck in the edge computing architecture. Even if enabled with dedicated hardware capabilities, these devices provide only reduced computing power which might not be sufficient for computing intensive applications like Deep Neural Networks (DNNs).

\begin{figure}
\begin{subfigure}{.25\textwidth}
  \centering
  \includegraphics[width=1\linewidth]{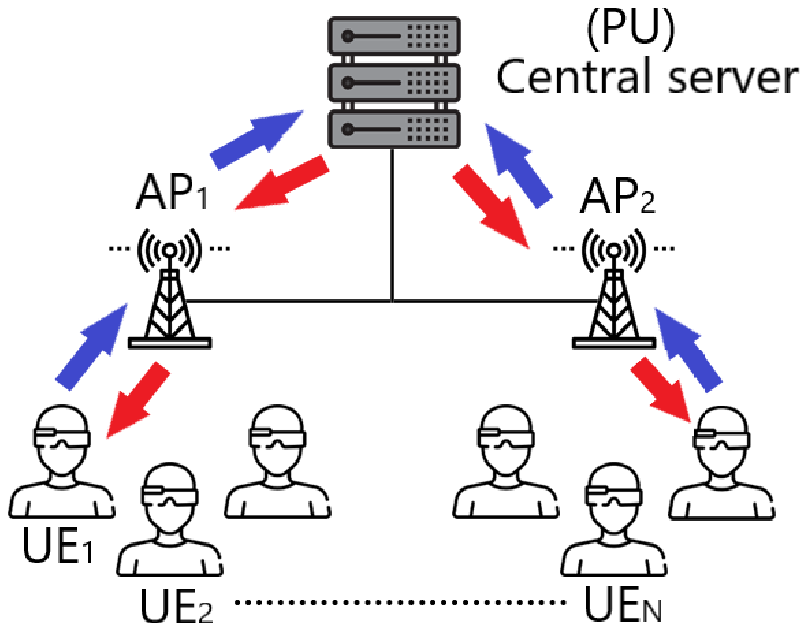}
  \caption{Centralized architecture}
  \label{fig:sfig0_1}
\end{subfigure}%
\begin{subfigure}{.25\textwidth}
  \centering
  \includegraphics[width=1\linewidth]{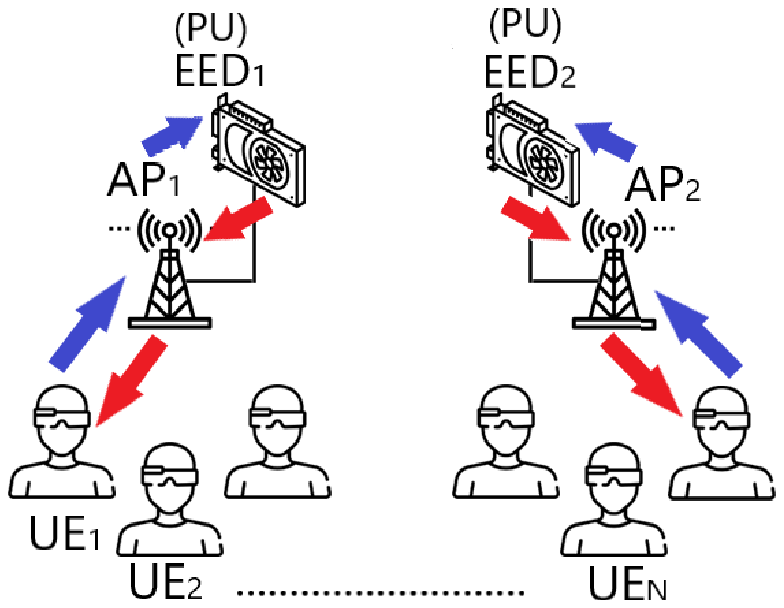}
  \caption{Distributed architecture}
  \label{fig:sfig0_2}
\end{subfigure}
\caption{The centralized vs. distributed architecture}
\label{fig:fig}
\end{figure}

In this paper we aim to answer the following question: \emph{can EEDs be a viable solution to support applications with stringent computing and responsiveness requirements?} To understand the adoptability of these systems, we compare two different deployment strategies for edge cloud deployments (Fig.\ref{fig:fig}): (a) a centralized approach that uses shared server grade equipment to support object recognition tasks and (b) a distributed system that exploits EEDs under the same application scenario. We focus our study on a realistic use case, in which User Equipment (UE) devices like Augmented Reality (AR) glasses or smartphones capture videos continuously and send frames to  some Processing Units (PUs), which can either be a single server grade hardware or a set of EEDs. In particular, we use object recognition models based on DNN as the target computing task.

Our contributions can be summarized as follows: 
\begin{enumerate}
    \item We develop a simple mathematical model (\S\ref{sec:analytical-model}) to study how varying the number of users affects the total performance of the system.
    \item To correctly parametrize such a model, we carry on a set of empirical measurements  (\S\ref{sec:measurement-campaign}) to evaluate the latency and the accuracy of a state deep learning algorithm for video analytics (e.g., object detection and recognition) on different platforms (i.e. Coral Dev Board, Jetson Nano, and GPU enabled server). 
\item Finally, we validate the analytical model through simulations (\S\ref{sec:simulation-setup}) in NS3 \cite{ns3} and conduct the performance comparison (\S\ref{sec:system-capacity-and-responsiveness}) between centralized and distributed solutions in a real life scenario. Our results show that, if we want to support a high number of users, we need to deploy a certain number of EEDs instead of a single central server with high computational capacities when it is the case of an AR application requiring low/mid responsiveness (Fig.~\ref{fig:ach} and~\ref{fig:achG}). For applications that demand high responsiveness, the only solution is the central server with high computational capacity, which anyways can support very few users.
\end{enumerate}

\section{System Description}
\label{sec:system-description}

We consider an AR application, whose aim is to capture videos of the environment, with \textbf{UEs} such as smart glasses or phones, and show to the user an appropriate description. This kind of application is envisaged in the context of Industry 4.0, where human operators are provided with information on their visual field, during maintenance or training. It can also be considered for cultural or leisure activities, as in smart museums, to identify artifacts.

A camera on the UE captures a video and continuously sends frames to processing units over the wireless link, via wireless \textbf{Access Points} (\textbf{APs}). The \textbf{Processing Units} (\textbf{PUs}) can be either a single \textbf{Central Server} or several \textbf{EEDs}, e.g., Jetson Nano or Google TPU Boards (Fig.~\ref{fig:fig}). 
PUs host a pre-trained DNN, which performs object detection and recognition. Finally the classification result (and optionally additional information) is sent back to the users from the PUs through the wireless link, and is shown in the field of view in the UE.

We contrast two architectures: (i) a \textbf{Centralized Architecture}, where the PU is one single central server, equipped with a powerful GPU and (ii) a \textbf{Distributed Architecture}, where PUs are EEDs, and we deploy one EED per AP. 

We want to show in which cases one is to be preferred to the other. We consider three kinds of applications (Table~\ref{tab:1}). They are \textbf{Low Responsiveness} (\textbf{LR}), \textbf{Mid Responsiveness} (\textbf{MR}) and \textbf{High Responsiveness} (\textbf{HR}) applications.
We also consider two types of wireless links, namely 802.11.ac \cite{80211ac} and 802.11.ax \cite{80211ax}, which we denote with \textbf{Low Wireless Capacity} and \textbf{High Wireless Capacity}.

\section{Related Work}
We now review the different solutions proposed in the literature to enable AR.
AR is demanding both in terms of complex computation and low latency. Indeed, computation is mostly performed by DNN and the result, e.g., classification, must be sent back to the User Equipment in few milliseconds (Table~\ref{tab:1}).
Recent work proposes optimizing DNNs in order to run them directly into the smartphones~\cite{Balan2017}. However, due to energy and computational constraints of smartphones, the latency is about 600 ms, which is too high for mid and high responsiveness applications (Table~\ref{tab:1}). For this reason, most of the interest has now moved to \emph{offloading} techniques: the computation takes place in a processing unit (PU), different than the UE. While this generally decreases the computation latencies, it also adds a network latency. In some work~\cite{Balakrishnan2015}, the PU is represented by the Cloud. However, the latency remains above 400 ms (Fig. 9-12 of~\cite{Balakrishnan2015}), due to the high network latency to reach the Cloud. Moreover, the bandwidth consumed on Metro Area Networks risks to be unmanageable~\cite{Satyanarayanan2015}.

The Edge Computing (EC) paradigm promises to keep network latency small, thanks to the proximity of PUs and UE.
An AR application with object detection, similar to our considered set-up, is studied in~\cite{Zhang2019}, where images from the User Equipment (UE) are matched against a database of images to find the most similar. Most of the computation is done by the edge servers, which also cache a part of the database. Only if the correct images are not found in such cache, the video frames are sent to the Cloud to be processed there.
According to the authors, their system can achieve 300ms end-to-end latency when the edge server is used by 36 users simultaneously.
The focus of~\cite{Q. Liu} is instead a load-balancing algorithm that chooses in which of the available edge servers the video frames from UE  should be sent.
A limit of \cite{Q. Liu} is that it needs to solve an optimization problem. Between one resolution and the next, requests reside in a queue, for up to 100 ms. Therefore, this queue-and-optimize approach introduces a non-negligible additional latency which does not fit all the AR applications requirements as shown in Table \ref{tab:1}. Moreover, they assume to use Edge servers, on average 50 ms away, which, alone, exceeds the most stringent requirements. In our work we tackle instead a wider range of application requirements and, in order to satisfy the most stringent, we do not apply any load-balancing optimization and we place, in our distributed architecture, one processing unit (PU) per access point (AP).
Liu et Al.~\cite{L. Liu} observe that, to adapt to the visual field on the UE, which continuously moving  in AR applications, the overall latency must be below 20 ms. To achieve such objective, they propose a set of solutions consisting in performing part of the computation directly in the UE, in order to reduce the amount of information to offload toward the PUs. In this paper, instead, we do not consider any of such optimizations: our UE just needs to capture and sends the video and the entire processing is done in the PUs, in order to save computation and energy burden on battery-limited UE.
A novelty of our work is that, while in all the previous work mentioned here, the PUs are fully-fledged servers, we instead consider a fully distributed architecture, where the processing is entirely done on small and cheap (\textasciitilde 150\$) embedded devices and contrast it with the use of a fully-fledged server as PU.
In an alternative architecture proposed by~\cite{Satyanarayanan2015}, PUs are Cloudlets, i.e., virtual machines deployed at the edge of the network, i.e., in base stations. However, cloudlets can only run on fully-fledged servers. We want to understand, instead, the feasibility of low latency applications only relying on edge AI embedded devices as a replacement of fully-fledged servers.

\begin{table}[t]
  \small 
  \centering
  \caption{Augmented Reality requirements}
  \label{tab:1}
  
  \begin{tabular}{C{4cm}C{3cm}}
    \hline
    AR requirements  & Latency\\
    \hline
     Low Responsiveness & 500 ms~\cite{Balan2017}\\
     
     Mid Responsiveness  & 100 ms\\
     
     High Responsiveness  & 16 ms \cite{L. Liu}\\
    \hline
  \end{tabular}
\end{table}

\section{Performance Evaluation}
In order to assess the performance of the Centralized vs. Distribution architectures, we first provide a simple \textbf{analytical model} (\S\ref{sec:analytical-model}) to obtain equations for the latency and the maximum number of supported users (more complex models~\cite{Sikdar2004} will be considered in future work). In order to give realistic values to the model parameters and to assess the classification accuracy of the DNNs on the PU, we perform a \textbf{measurement campaign} (\S\ref{sec:analytical-model}) on a central server and two types of EEDs available in the market. We then validate our model in \textbf{simulation} (\S\ref{sec:simulation-setup}). We finally find the system capacity and latency (\S\ref{sec:system-capacity-and-responsiveness}).

\subsection{Analytical Model}
\label{sec:analytical-model}

Let us consider a network with $N$ users and a processing unit (PU).
User Equipment (UE) transmits a sequence of frames to one wireless Access Point (AP), which will then route it to the PU.
For our analytic model we resort to~\cite{Q. Liu}, with the additional simplifying assumption that we do not consider the latency of sending-back the detection and recognition results, as in \cite{Letaief2016}, as we assume they are mainly textual. Note that, as in~\cite{Q. Liu}, we do not consider the core network latency to transmit the frames from the AP to the PU, as it is negligible in practical scenarios (lower than 2 ms in~\cite{Li2017}). As in~\cite{Q. Liu}, we also ignore the effect of congestion in the AP-PU network segment.
Therefore, the system latency experienced by a user can be defined as :
\begin{align}
    \label{eq:system-latency}
    L = L^w + L^p
\end{align}

where $L^w$ is the wireless latency incurred by sending a video frame up to the AP and $L^p$ is the processing latency for object detection and recognition in the PU.
We denote with $L^\text{required}$ the latency required by the application (Table~\ref{tab:1}). Obviously, $L\le L^\text{required}$ must be guaranteed.

\subsubsection{Wireless Channel characterization}
The wireless latency is determined by the user’s video frame resolutions and wireless data rates. 
We denote with $R$ the goodput, i.e., the number of  TCP payload bits sent in a second.  It has been observed (\S~11.5 of~\cite{Bonald2011}) that when using TCP over a wireless network, the goodput $R$ is approximately insensitive to the number of stations. This is why we keep $R$ constant with respect of the number of UE devices . Moreover, a remarkable property of TCP flows on a wireless link is that data-rate is approximately equally shared among the N users (\S~11.5 of~\cite{Bonald2011}).
We can adopt such approximations, since we assume that UEs are relatively close to the AP and there are no obstacles. 
Therefore, the data-rate of a single UE is $r= \frac{R}{N}$.

A frame can be encoded in different resolutions $k\in\mathcal{K}$. Each resolution corresponds to $s_k^2$ pixels. The color depth $\sigma$ is the number of bits employed to encode one pixel. Therefore, the size of a frame is $D_k=\sigma\cdot s_k^2$, in bits and to transmit a frame from the UE to the AP, the wireless latency is is~\cite{Josilo2018}:
\begin{equation}
    \label{eq:wireless-latency}
    L^w = \frac{D_k}{r} = \frac{D_k\cdot N}{R} = \frac{\sigma\cdot s_k^2\cdot N}{R} 
\end{equation}

\subsubsection{Framerate and maximum number of users}
Let us denote with $f$ the framerate, i.e. the number of frames transmitted per second from the UE to the PU. Observe that $f$ does not necessarily correspond to the framerate at which the video is captured, which can indeed be larger than $f$. Indeed, for some application, we may choose to subsample the video frames and send to the PU only a fraction $p\le 1$ of the video frames (see \S3.1 of~\cite{Balakrishnan2015}). Note also that the deep learning strategies considered in this paper and the related literature we refer to here, take entire frames as input, independent of the video encoding. For instance, if video is encoded in MPEG where there is only one complete frame (I-frame) in a Group of Picture (GoP), since DNNs are only capable of processing I-frames, $f$ will refer to the number of I-frames sent to the PU, independent of the GoP and framerate of the video capture.

If we want to have a response within $L^\text{required}$ to events detected in the environment, we need to send at least a frame each $L^\text{required}$. Therefore, $f\ge 1/L^\text{required}$.
When $N$ users send $D_k$ bits in a shared wireless channel per each of their frames, then $f \cdot N\cdot D_k \le R$ and thus 
$N \le
    \frac{R}{D_k\cdot f}\le
    \frac{R\cdot L^\text{required}}{D_k}$.
Therefore, the number of users that can be supported simultaneously is:
\begin{align}
    \label{eq:n-max}
    N_{\text{max}} =
    \frac{R\cdot L^\text{required}}{D_k}
\end{align}

\subsubsection{Computational latency}
\label{sec:computational-latency}
As in~\cite{Q. Liu}, the computational latency to process a single frame is:
\begin{align}
    \label{eq:c27}
    L^p = \frac{c_k}{F_\text{pu}} \cdot N
\end{align}
where $c_k$ denotes the \emph{complexity} of the inference for the frame resolution $k$, $F_\text{pu}$ the amount of computation resources on the processing unit $\textit{pu}$ (which can be either a central server or a EED, as Jetson Nano or Google TPU board) and $N$ the number of users sharing the PU. $c_k$ depends on the frame resolution $s_k^2$ following a function $c_k = \psi(s_k)$. For simplicity, we set $\psi_\text{pu}(s_k^2)\triangleq\frac{\psi(s_k^2)}{F_\text{pu}}$ and rewrite~\eqref{eq:c27} as:
\begin{align}
    \label{eq:c27-bis}
    L^p = \psi_\text{pu}(s_k^2) \cdot N
\end{align}

We experimentally measure in \S\ref{sec:measurement-campaign} and Fig.~\ref{fig:inf_exp} the value of the function $\psi(s_k^2)$, for all resolutions $k\in\mathcal{K}$ and for the three types of PUs considered (Jetson Nano, Coral Dev, Central Server).

\subsection{Measurement Campaign}
\label{sec:measurement-campaign}

In order to make our analytical model operational for our comparison of Centralized vs. Distributed architectures, we perform a set of measurements on real devices, in order to fix the value of the resolution $k$, the bits-per-pixel $\sigma$ and find the complexity $c_k$.
Since we want our work to be general enough and applicable to different scenarios, where different objects may need to be detected, we use the Common Objects in COntext (COCO) dataset \cite{coco}.

As for the Centralized Architecture, our Central Server is equipped with an Intel(R) Xeon(R) CPU E5-2620 v3 @ 2.40GHz CPU, 32 GB of RAM and NVIDIA GeForce GTX980 GPU.
As for the distributed architecuture, instead,we experimented with two different EEDs: Jetson Nano Developer kit from Nvidia and Coral Dev Board by Google. 

We deploy on the PUs a pre-trained DNN called SSD mobilenet v2 \cite{mobnetv2}, which combines SSD for object detection and Mobilenet to classify the detected objects.
In order to run such a DNN on Coral Dev and Jetson Nano in an effective way, we convert the regular model to a lighter version using Tensorflow Lite Converter and TensorRT, respectively, with their default configurations.
Converting the regular model reduces its file size and introduces optimizations suitable for EEDs.
The TensorFlow Lite Converter takes a trained TensorFlow (TensorFlow only) model as input and outputs a TFLite (.tflite) file, a FlatBuffer-based file containing a reduced, binary representation of the original model.
On the other hand, TensorRT, built on CUDA, NVIDIA’s parallel programming model, enables to optimize inference for all deep learning frameworks by fusing specific Layers and Tensors in one layer which improves the use of GPU memory and bandwidth.

All the measurements of this section are obtained with a color depth $\sigma=8$ bits/pixel. In fact, we observe that increasing it does not considerably improve accuracy.

The frame resolution $k$ chosen is one of the most important parameters: on the one hand, increasing it improves the classification accuracy, on the other hand it increases the amount of bits to send over the wireless link. Therefore, a good trade-off must be found.

In this paper, we consider 6 frame resolutions: $\mathcal{K}=\{$100x100, 200x200, 300x300, 500x500, 800x800 and 1000x1000$\}$ pixels. For each frame resolution, we perform object detection+classification inference on 200 images.

\begin{figure}
\begin{subfigure}{.25\textwidth}
  \centering
  \includegraphics[width=1\linewidth]{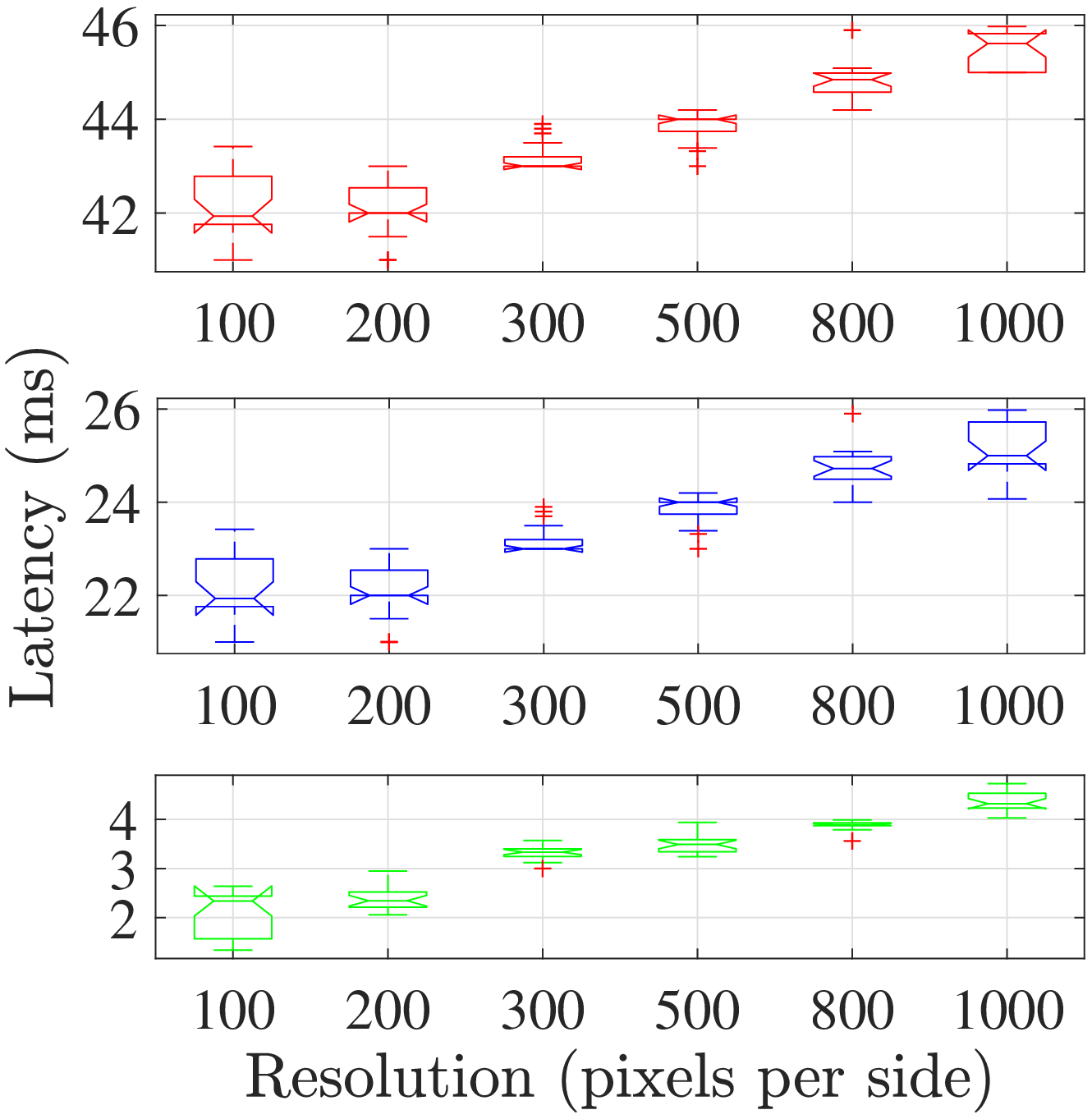}
  \caption{Inference time. }
  \label{fig:inf_exp}
\end{subfigure}%
\begin{subfigure}{.25\textwidth}
  \centering
  \includegraphics[width=1\linewidth]{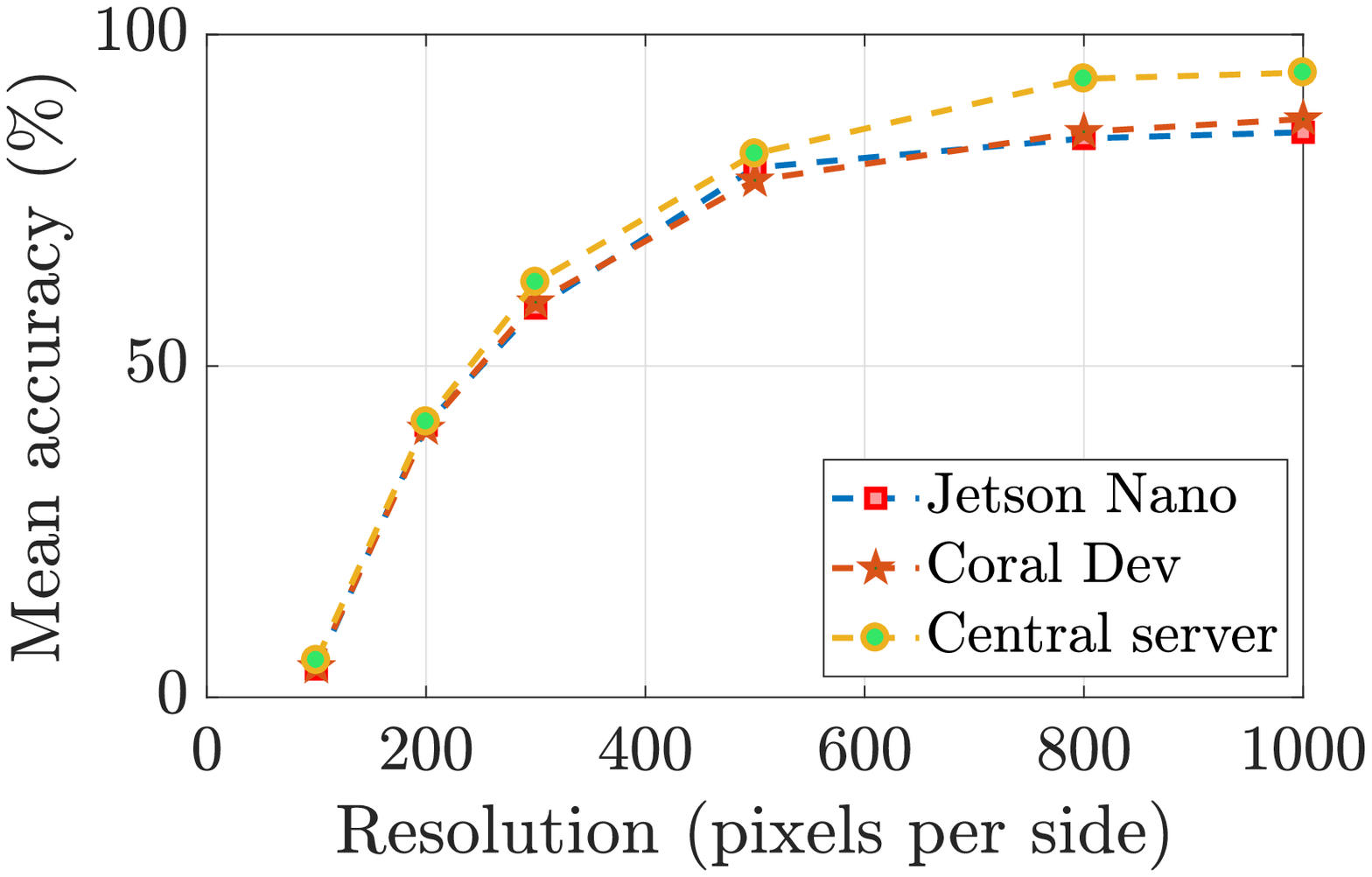}
  \caption{Detection mean accuracy}
  \label{fig:accuracy}
\end{subfigure}
\caption{Experiments results on different platforms}
\label{fig:experiments}
\end{figure}

Fig.~\ref{fig:inf_exp} shows the inference latency, i.e., the time spent by the DNN between the instant it receives the frame up to the instant in which it provides the classification. As in~\cite{Q. Liu}, we fit the curve $\psi_\text{pu}(s_k^2)$ with a cubic regression, i.e., $\psi_\text{pu}(s_k^2)=a + b s_k^3$. The values of the coefficients $a,b$ for with the types of PUs considered are given in Table \ref{tab:2}. 
The resolution has a negligible impact on the inference time (which is confirmed by the small $b$ coefficients). It is instead very evident that the inference time greatly depends on the type of PU. The Central server is one magnitude faster than Jetson Nano, which remarkably slower than Coral Dev. The goal of this paper is however not to compare different Edge AI platforms. Here we want just to show that it is important to take into account the differences between them, and always starts the study of the performance in such scenarios with a measurement campaign on the real devices considered.

\begin{table}[t]
  \small 
  \centering
  \caption{Inference time fitted curve parameters}
  \label{tab:2}
  
  \begin{tabular}{ccc}
    \hline
    Type of PU  & $a$ & $b$ \\
    \hline
    Central server & 3.23 & 9.56 $e-10$\\
    
    Coral Dev Board & 20.98 & 3.37 $e-09$\\
     
    Jetson Nano  & 41.10 & 7.15 $e-09$\\
     
    \hline
  \end{tabular}
\end{table}

Fig.~\ref{fig:accuracy} shows that, in order to achieve a satisfying accuracy, we need a resolution larger than $500$p. For this reason, we fix in what follows $k=600p$. Note also that accuracy with the Central Server is better than with EDDs, which is due to the optimizations applied when converting the DNN to the format suitable for EEDs.

Having fixed $k, \sigma$ and $c_k=\psi_\text{pu}(s_k^2)$, for all the PUs under consideration, we can now use the analytical model of \S\ref{sec:analytical-model} to obtain numerical values for the wireless latency~\eqref{eq:wireless-latency}, computational latency~\eqref{eq:c27-bis} and the maximum number of supported users~\eqref{eq:n-max}.

Note that our measured accuracy matches the one observed in~\cite{Q. Liu}, while results are quite different in what concerns the inference time, which can be explained by the different Edge AI platform considered there. This confirms that a solid performance evaluation of Edge AI solutions should always start from a preliminary measurement campaign on the real devices, as we do here.

\subsection{Simulation setup}
\label{sec:simulation-setup}
We now validate the analytical model via simulation in NS-3. We model Low and High Wireless capacity with the NS3 implementation of 802.11ac in MIMO 3x4 mode with a 80 Mhz channel and 802.11ax in MIMO 3x4 mode with a 160 Mhz channel width. The goodput is $R=450$Mbps and $R=1$Gbps, respectively (Fig.3 of~\cite{Kurt2015} and Table 1 of~\cite{intro2020}).

\begin{figure}
\begin{subfigure}{.25\textwidth}
  \centering
  \includegraphics[width=1\linewidth]{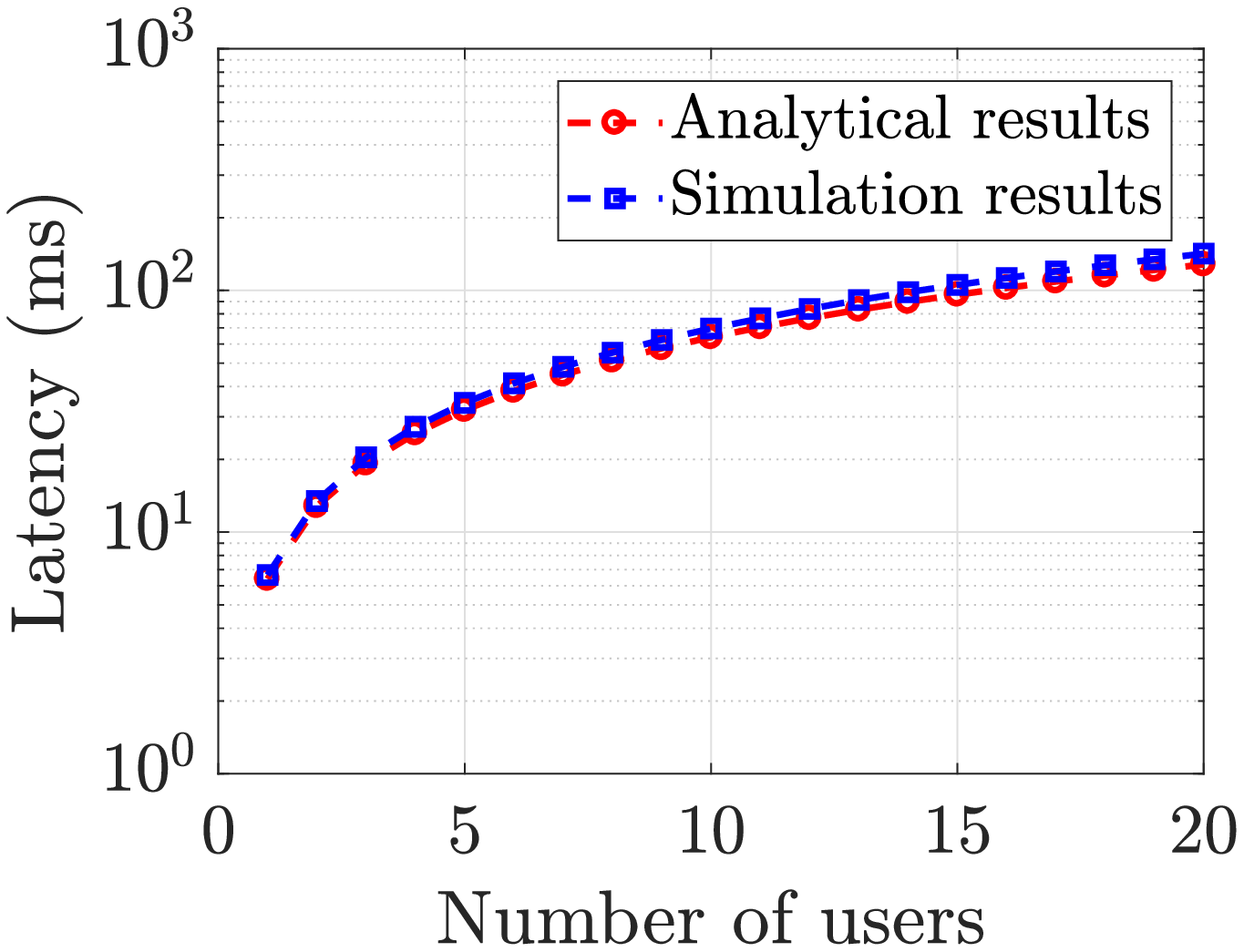}
  \caption{Wireless performance validation}
  \label{fig:sfig1_1}
\end{subfigure}%
\begin{subfigure}{.25\textwidth}
  \centering
  \includegraphics[width=1\linewidth]{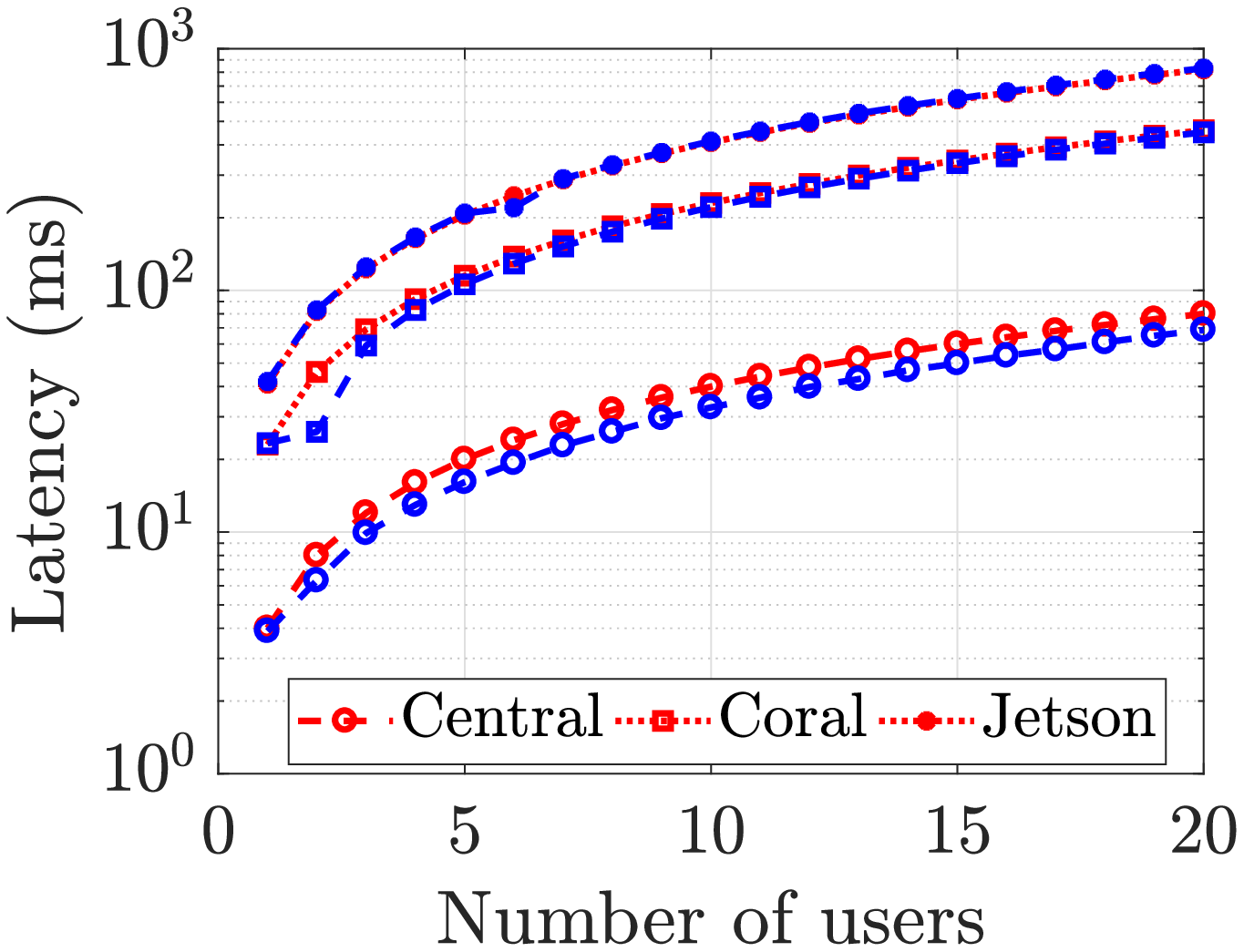}
  \caption{Computation validation}
  \label{fig:sfig1_2}
\end{subfigure}
\caption{Validation of the analytical model}
\label{fig:validation}
\end{figure}

We first focus on one single AP and $N$ UEs connected to it, in both the Centralized and Distributed scenario, with Low Wireless Capacity. We defer to \S\ref{sec:system-capacity-and-responsiveness} the case of multiple APs and High Wireless Capacity. We model the UEs, the AP and the PUs server with the NS-3 class \texttt{NodeContainers} whose individual size corresponds to the number of elements from each class (EEDs, APs and PUs). 
We measure the wireless latency plus the inference time, varying the number of UEs.
Since NS-3 does not model the computational capacity of a server, we model a PU as a queue. This queue has a constant serving time which corresponds, for each platform, to the inference latency for one frame, as measured in \S\ref{sec:measurement-campaign}.
Since the timescales we are considering are very short (Table~\ref{tab:1}), much faster than human movement, it is reasonable to simulate, every time, a static snapshot of the system, in which users do not switch from an AP to another. Our UEs send continuously frames to the PU, sending the next frame immediately after receiving the response. The simulation time is 17 minutes. Fig.~\ref{fig:validation} shows that simulation results match the analytical ones.

We then simulate two real scenarios where we have multiple users in the coverage of multiple APs. As in Fig.~\ref{fig:fig}, for the distributed architecture, we deploy an EED at every AP (in a way that we have one EED per AP) while for the centralized architecture we deploy a single server connected to all of the APs. In each scenario, we let the total number of users vary, as well as the number of APs (the more the APs, the less the users-per-AP). Note that the number of APs corresponds, in the simplest case, to the number of rooms a the museum/exposition or industrial plant.

\subsection{System Capacity and Latency}
\label{sec:system-capacity-and-responsiveness}

\begin{figure}
\begin{subfigure}{.25\textwidth}
  \centering
  \includegraphics[width=1\linewidth]{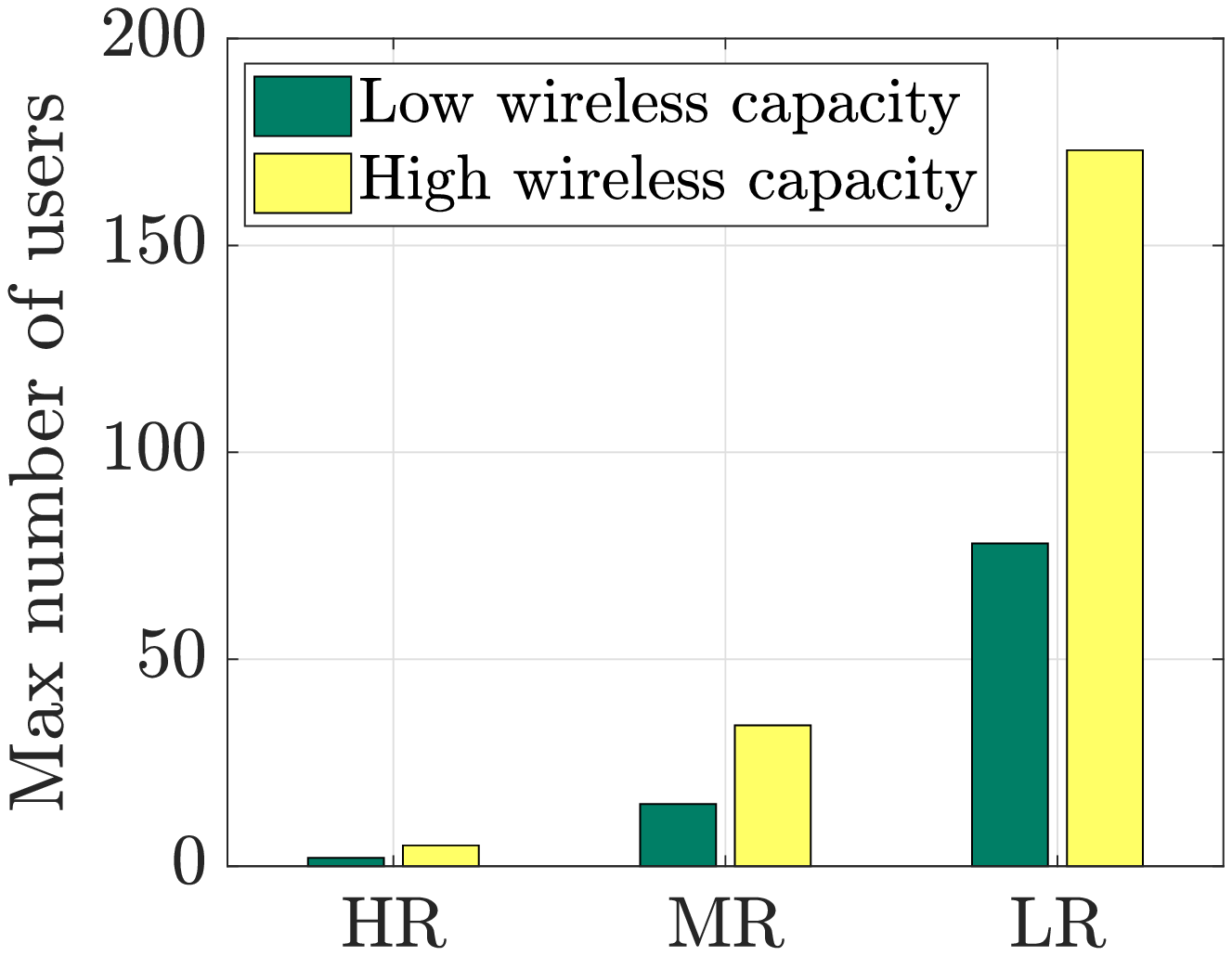}
  \caption{Maximum number of users}
  \label{fig:anal1}
\end{subfigure}%
\begin{subfigure}{.25\textwidth}
  \centering
  \includegraphics[width=1\linewidth]{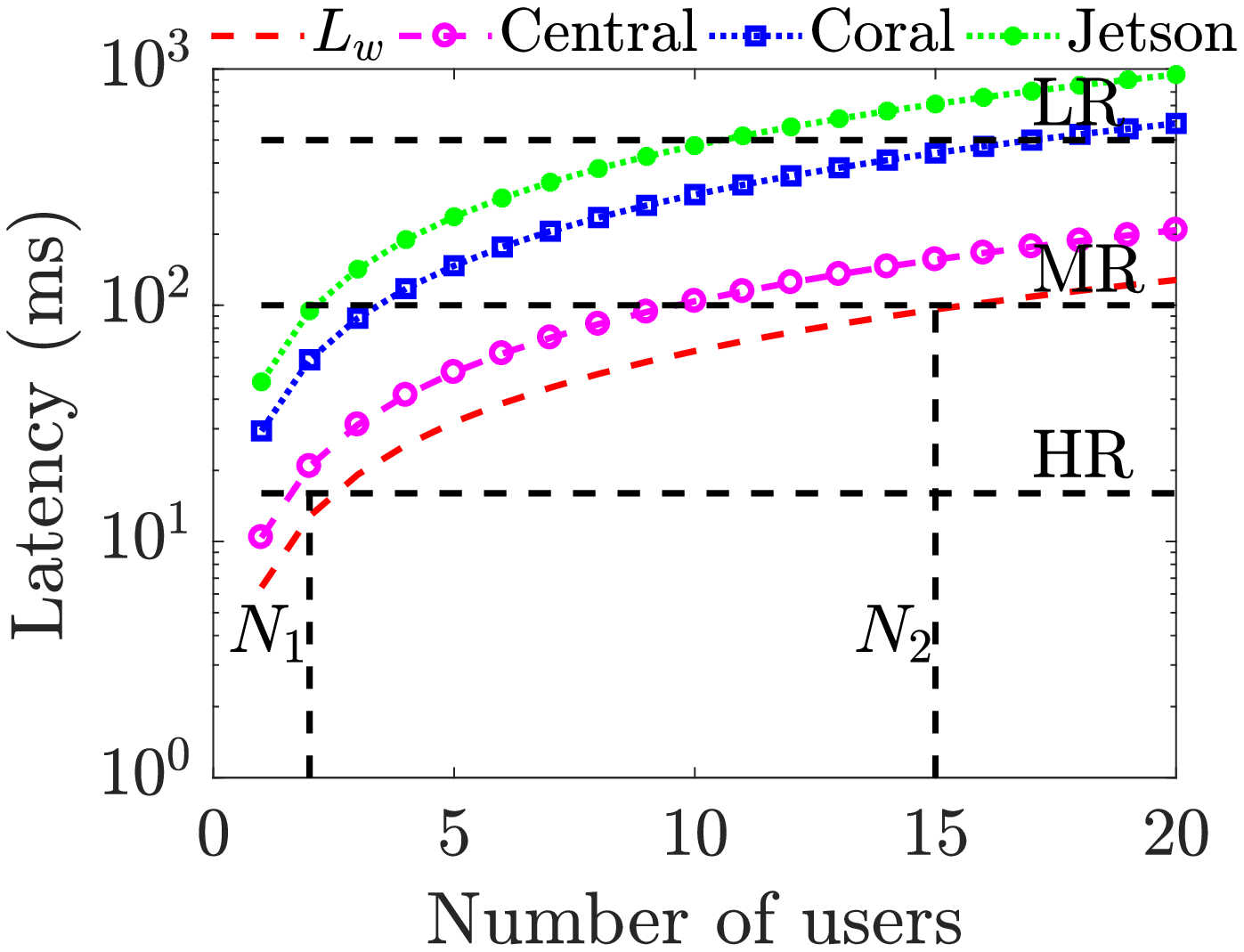}
  \caption{System Latency}
  \label{fig:anal2}
\end{subfigure}
\caption{System capacity and latency.} 
\label{fig:analy}
\end{figure}

The capability of sustaining AR applications depends on two factors: the system capacity, i.e., the number of simultaneous users that can be supported, and the system latency, i.e., how fast can each single request be satisfied.
The system capacity, on its turn, depends on two elements: the number of users supported by the wireless channel and the processing capability of the PUs. In what follows, we first study the limitations imposed by the wireless channel (\S\ref{sec:wireless-channel}), then the system latency (\S~\ref{sec:system-latency}). Considering system capacity and latency together, we finally study which requirements can be supported by the centralized and distributed architecture (\S~\ref{sec:supported-applications}).

\subsubsection{Limitations imposed by the wireless channel}
\label{sec:wireless-channel}
The maximum number of users $N_\text{max}$ per AP supported by the wireless channel can be obtained by~\eqref{eq:n-max}. 
Fig.~\ref{fig:anal1} shows $N_\text{max}$ with Low and High wireless capacity technologies, namely providing $R=450$ Mbps and $R=1$ Gbps and with High, Mid and Low responsiveness requirements (HR, MR, LR), i.e., $L^\text{required}=\{16, 100, 500\}$~ms (see Table~\ref{tab:1}). Observe that $N_\text{max}$ only takes into account the limitation imposed by the wireless channel and not by the Processing Units (PUs). This limitation appears the same, in both distributed and centralized architectures. In other words, $N_\text{max}$ indicates the maximum number of supported users in presence of ideal PUs of infinite capacity.

Fig.~\ref{fig:anal1} shows that the number of users that can be supported by the wireless channel is very low when we need to send a frame every 16 ms (HR) e.g. 2 users with the low wireless capacity and 5 users with the high wireless capacity. The maximum number of users supported by the wireless channel goes to 15 and 34 for the low wireless capacity and the high wireless capacity respectively when we need to send a frame each 100 ms (MR). When it comes to LR (sending a frame every 500 ms), the wireless channel can support more than 75 users for low wireless capacity and more than 150 users for the higher wireless capacity. 

\subsubsection{System Latency}
\label{sec:system-latency}

The system latency (see~\eqref{eq:system-latency}) computed by the analytical model is depicted in Fig.~\ref{fig:anal2}, where we focus on just one AP and we assume that there is one PU, namely a Coral Dev or Jetson Nano or Central server, behind the AP. We also assume Low wireless capacity $R=450$ Mbps.
Note that these results are a lower bound for the centralized architecture, in which there is one PU only for all the APs (see Fig.~\ref{fig:fig}). $N_1$ and $N_2$ in Fig.~\ref{fig:anal2} refer to the maximum number of users (see~\eqref{eq:n-max}) supported by the wireless channel for High Responsiveness (HR) and Mid Responsiveness (MR) applications, respectively (see Table \ref{tab:1}). For Low Responsiveness (LR) applications, instead, $N_\text{max}>20$.

Fig.~\ref{fig:anal2} shows that already by considering the wireless latency $L^w$ only, with more than 2 users, it exceeds the requirements of HR. Adding also the processing time $L^p$, only one user is supported in HR (and only with the centralized server).
 However, we can achieve Mid Responsiveness and Low Responsiveness requirements with the distributed architecture but with a limited number of users compared to the centralized architecture.

\subsubsection{Supported Applications}
\label{sec:supported-applications}

\begin{figure}
\begin{subfigure}{.25\textwidth}
  \centering
  \includegraphics[width=1\linewidth]{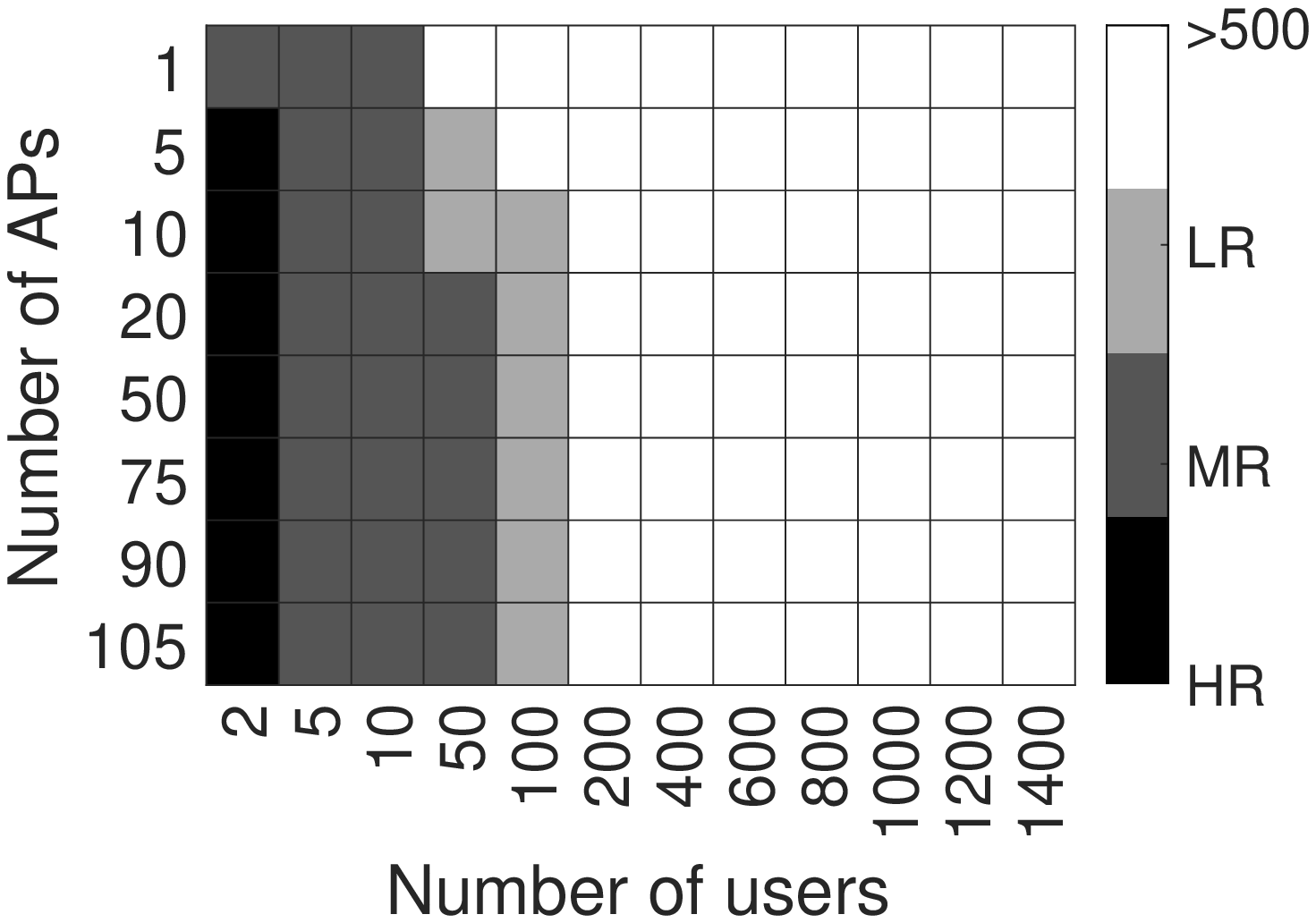}
  \caption{Centralized architecture}
  \label{fig:ach1}
\end{subfigure}%
\begin{subfigure}{.25\textwidth}
  \centering
  \includegraphics[width=1\linewidth]{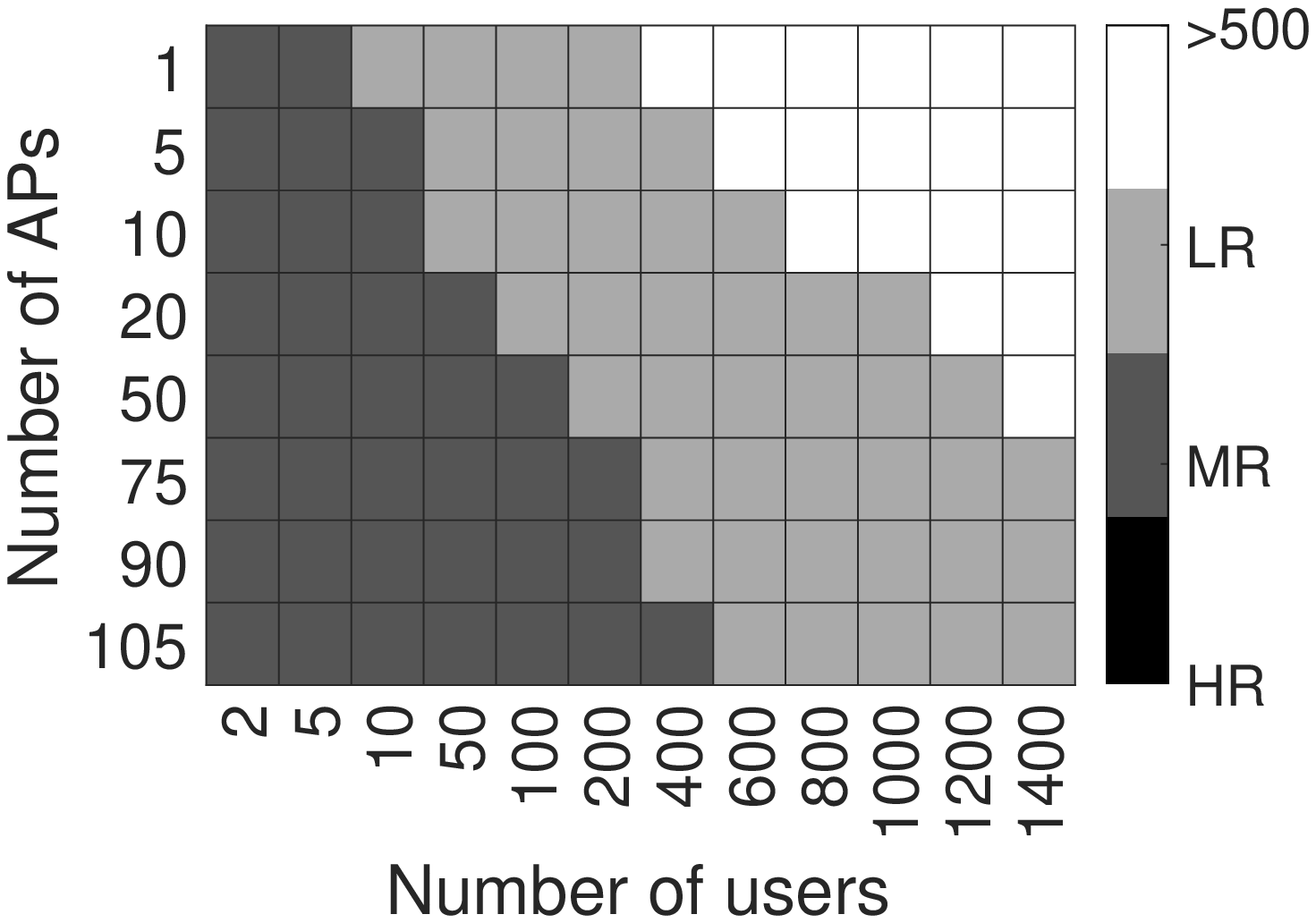}
  \caption{Distributed architecture}
  \label{fig:ach2}
\end{subfigure}
\caption{Achievable requirements for $R =$ 450 Mbps.}
\label{fig:ach}
\end{figure}

\begin{figure}
\begin{subfigure}{.25\textwidth}
  \centering
  \includegraphics[width=1\linewidth]{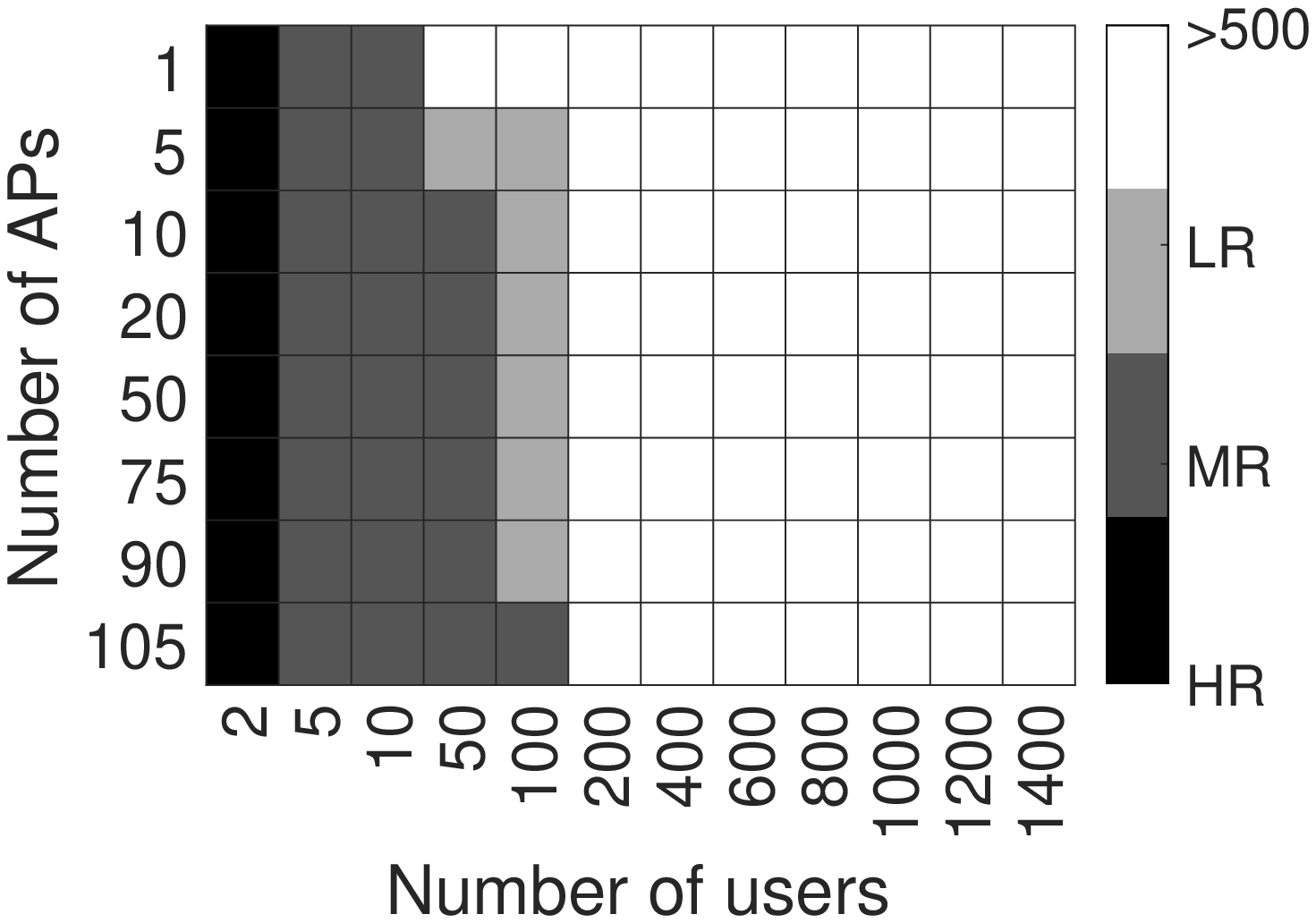}
  \caption{Centralized architecture}
  \label{fig:ach11}
\end{subfigure}%
\begin{subfigure}{.25\textwidth}
  \centering
  \includegraphics[width=1\linewidth]{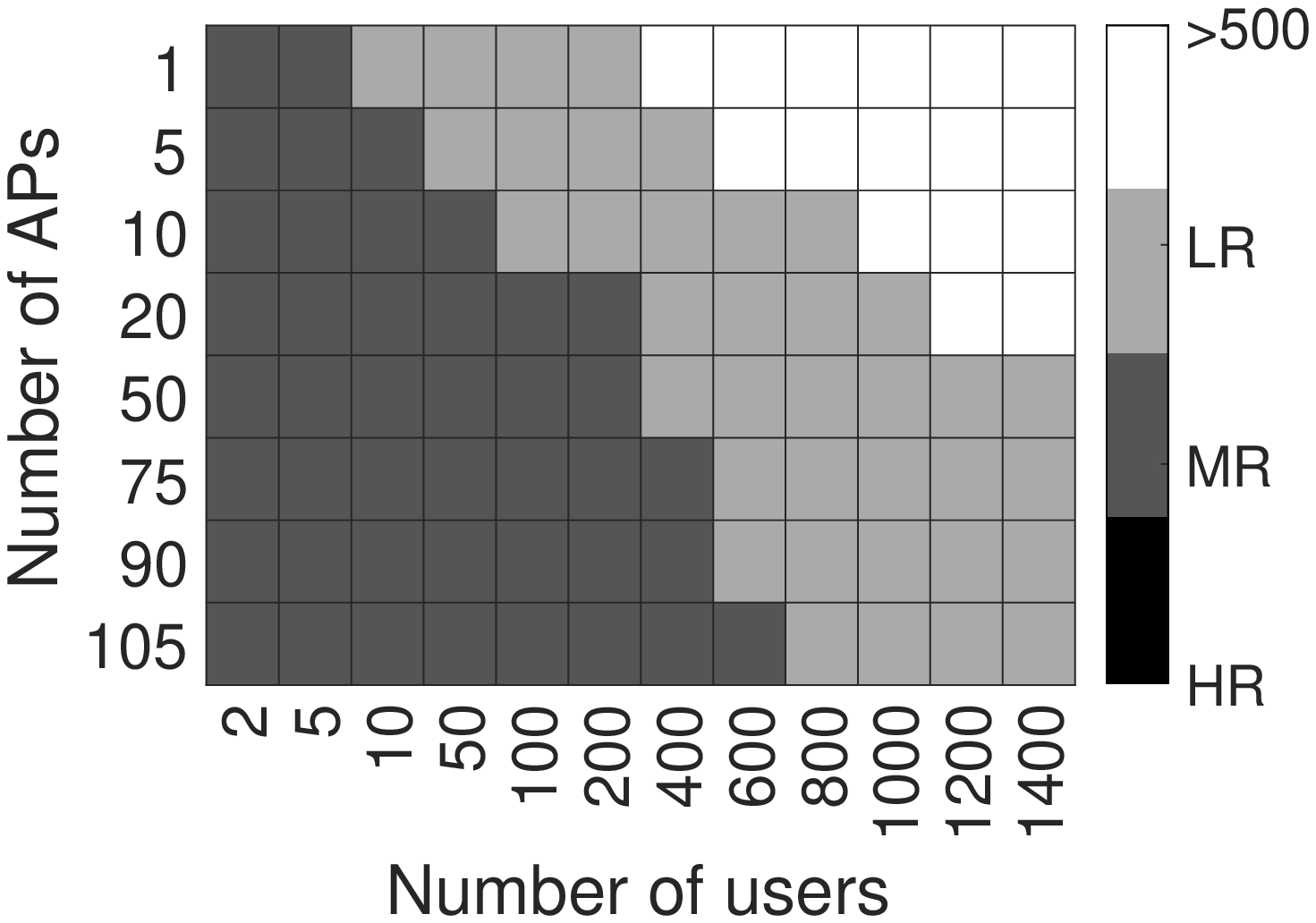}
  \caption{Distributed architecture}
  \label{fig:ach12}
\end{subfigure}
\caption{Achievable requirements for $R =$ 1 Gbps.}
\label{fig:achG}
\end{figure}

In Fig.~\ref{fig:ach} and~\ref{fig:achG}, we show the requirements achievable (Table \ref{tab:1}) in 4 scenarios, assuming Low and High wireless data rates (\S\ref{sec:system-description}) and the distributed vs. centralized architectures (Fig.~\ref{fig:fig}). We consider a range of number of users $\#U$ going from 2 users up to 1400 users, which is the number of simultaneous visitors in a big museum as the Louvre. Such users are distributed across different APs, whose number $\#AP$ is varied between 1 and 105. The number of users sharing the wireless channel of a single AP is $N=\lceil\frac{\#U}{\#AP}\rceil$, i.e., the minimum integer bigger than $\frac{\#U}{\#AP}$. 

 A certain requirement $L^\text{required}$ is satisfied if $N\le N_\text{max}$ (see~\eqref{eq:n-max}) and $L=L^w+L^p\le L^\text{required}$. We check such conditions via our analytical model (\S\ref{sec:analytical-model}). In Fig.~\ref{fig:ach} and~\ref{fig:achG} we represent which of the requirements of Table~\ref{tab:1} is satisfied.
 The results presented in Fig.~\ref{fig:ach} and~\ref{fig:achG} are obtained via NS3 simulation. Furthermore, we verified that they are also matched by the analytical model, which lets us compute the wireless latency $L^w$, based on~\eqref{eq:wireless-latency}, and the processing latency via~\eqref{eq:c27}:
 \begin{small}
\begin{align}
\label{eq:Lp}
L^p=
\begin{cases}
\frac{c_k}{F_\text{pu}}\cdot \#U    & \text{for the centralized architecture} \\
\frac{c_k}{F_\text{pu}}\cdot N    & \text{for the distributed architecture} 
\end{cases}
\end{align}
 \end{small}
 
 The results show that HR applications are only supported by the centralized architecture and only with few users. In all the other situations, the distributed architecture outperforms the centralized and can support MR and LR applications with a relatively large number of users, which is instead infeasible for the centralized architecture. Also observe that, as expected, increasing the number $\#AP$ of access points allows to support more demanding applications, as it reduces contention in the wireless channel in the centralized architecture. This also holds for the distributed architecture, in which the benefit of increasing $\#AP$, and thus also the PUs (see Fig.~\ref{fig:fig}), are more evident, since it also reduces contention in processing (see~\eqref{eq:Lp}). Also note that adopting the latest wireless technology with high capacity helps to support more demanding applications.

\section{Conclusion and Future Work}
This paper presents a comparison of classic centralized systems vs. distributed systems based on Embedded Edge Devices (EEDs) for deploying AR applications with different requirements. We develop an analytical model to represent system capacity and latency of the two alternative systems.
We evaluated the performance of deep learning algorithms for video analytics (e.g., object detection and recognition) on such devices via a measurement campaign. We parametrized our analytical model based on such measurements. We developed a NS3 simulator and verify that its results match the analytical model. The analytical model allowed us to study the performance with multiple users up to the order of a thousand. 

We show in this paper that for a certain AR applications with high responsiveness requirements, the only solution is a powerful central server. Only few users can be supported. On the other hand, other AR applications with less stringent demands (LR and MR) can be well supported by a distributed architecture based on EEDs. More importantly, only the distributed architecture is able to support a large number of users.

In future work, we will introduce the monetary cost into the comparison of centralized vs. distributed architecture, based on the products available on the market. We will optimize the EED assignment to the AR user by introducing on-line strategies, which take decisions at every new user. Focusing on improving the power consumption management, the challenge is to develop strategies in order to optimize transmission power and offloading data. Another challenge is to explore the performance of other DNN architectures and probably combinations between CNNs and Recurrent Neural Networks.
\addtolength{\textheight}{-12cm}   % This command serves to balance the column lengths

%%%%%%%%%%%%%%%%%%%%%%%%%%%%%%%%%%%%%%%%%%%%%%%%%%%%%%%%%%%%%%%%%%%%%%%%%%%%%%%%

\end{document}